\documentclass[aps,prb,twocolumn,superscriptaddress,showpacs,floatfix,reprint]{revtex4-1}

\usepackage{amsmath}
\usepackage{amssymb}
\usepackage{graphicx}

\newcounter{fignum}

\newcommand{\mbf}{\mathbf}

\begin{document}


\title{Early stages of magnetization relaxation in superconductors}

\author{Mihajlo Vanevi\' c}%
\affiliation{Department of Physics, University of
Belgrade, Studentski trg 12, 11158 Belgrade, Serbia}

\author{Zoran Radovi\' c}%
\affiliation{Department of Physics, University of
Belgrade, Studentski trg 12, 11158 Belgrade, Serbia}

\author{Vladimir G. Kogan}%
\affiliation{Ames Laboratory DOE, Ames, Iowa 50011, USA}

\begin{abstract}
Magnetic flux dynamics in type-II superconductors is studied
within the model of a viscous nonlinear diffusion of vortices for
various sample geometries. We find that time dependence of
magnetic moment relaxation after the field is switched off
can be accurately approximated by
$m(t)\propto 1-\sqrt{t/\tilde\tau}$ in the narrow initial time
interval and by $m(t)\propto (1+t/\tau)^{-1}$ at later times
before the flux creep sets in. The characteristic times
$\tilde\tau$ and $\tau$ are proportional to the
viscous drag coefficient $\eta$. Quantitative agreement with
available experimental data is obtained for both conventional and
high-temperature superconductors with $\eta$ exceeding
by many orders of magnitude the Bardeen-Stephen coefficient for free vortices.
Huge enhancement of the drag, as well as its exponential temperature
dependence, indicates a strong influence of pinning centers on the flux diffusion.
Notwithstanding the complexity of the vortex motion in the presence of
pinning and thermal agitation, we argue that the initial relaxation of
magnetization can still be considered as a viscous flux flow with an effective
drag coefficient.
\end{abstract}



\date{\today}

\maketitle

Magnetic flux penetrates a type-II superconductor in the form
of discrete quantized vortices. Vortex structures in
conventional and high-temperature superconductors display
remarkable complexity both in equilibrium\cite{CrabtreeNelsonPHYSTODAY50-97,BrandtRPROGPHYS58-95}
and dynamic regimes.%
\cite{AndersonKimRMP36-64,bardeen_theory_1965,beasley_flux_1969,%
TinkhamBook,%
blatter_vortices_1994,yeshurun_magnetic_1996,%
marchetti_hydrodynamics_1990,%
okada_PRB86-12,RaesMoshchalkovPRB86-12,lin_vortex_2012}
Relaxation of the magnetic moment of superconductors is achieved through
initial viscous flux flow
\cite{kunchur-PRL70-93,moshchalkov_early_1989,*MoshchalkovPHYSICAB169-91,%
pardo_observation_1998,troyanovski_collective_1999,
deng_relaxation_2012,monarkha_violation_2012}
and slow, logarithmic in time, thermally activated creep.%
\cite{feigelman_theory_1989,vinokur_exact_1991,abulafia_local_1995,ProzorovPRB58-98,gurevich_time_1993}
Thermally-assisted hopping of vortices and vortex bundles between local minima
in the random pinning potential is characteristic of both the creep and the
flux flow under a driving force. In the latter, the hopping gives rise to the
viscous drag coefficient $\eta \propto e^{U/kT}$, where $U$
is the effective activation energy and $T$ is the temperature.\cite{abulafia_local_1995}
A free flux flow regime can be realized at microwave frequencies
($10-100$GHz) when the effect of the pinning is negligible.
Measurements of surface impedance give viscous drag coefficients
$\eta_0\sim 10^{-6} - 10^{-7}\,\rm{Ns}/\rm{m}^2$ at low temperatures for
all superconductors, e.g., conventional
$\rm{NbSe}_2$,\cite{HorCrabtreeAPL87-05}
cuprates YBCO, BSCO,\cite{pompeo_reliable_2008,GolosovskyDavidovSUPERCONDSCITECHNOL9-96}
and pnictide LiFeAs.\cite{okada_PRB86-12}
The order of magnitude is in accordance with the
Bardeen-Stephen result for the viscous drag,
$\eta_0=\Phi_0 H_{c2}/\rho_n c^2$,
caused by dissipation in the vortex core
($\Phi_0=hc/2e$ is the flux quantum, $\rho_n$ is the normal-state resistivity,
and $H_{c2}$ is the upper critical field).\cite{bardeen_theory_1965}

In this paper we study early stages of the flux dynamics after
switching off the external magnetic field. We use a simple hydrodynamic
approach:
The local force the vortex experiences due to
interaction with other vortices, the surface,
and the local quenched disorder (pinning centers) is described
by an effective viscosity $\eta \gg \eta_0$.
The same approach successfully describes the vortex creep,
if supplemented by a phenomenological model of
current-dependent or time-dependent activation energy,
$U=U_c \ln(j_c/j)$ or $U=kT\ln(t/t_0)$,
where $j_c$ is the critical current and $t_0$ is the characteristic
time scale for flux creep.%
\cite{feigelman_theory_1989,vinokur_exact_1991,abulafia_local_1995,ProzorovPRB58-98,gurevich_time_1993}

We consider a model of massless vortex
motion where the driving Lorentz force equals the viscous drag
$(1/c)\mbf J \times \mbf \Phi_0 - \eta\mbf v =0$. Here, $\mbf J$
is the current density, $\mbf v$ is the vortex velocity, and
$\eta$ is a viscous drag coefficient.
For magnetic induction $\mbf B = n\mbf \Phi_0$
related to the vortex density $n$,
the force balance equation reads
$(1/c)\mbf J \times \mbf B - \eta |B|\mbf v/\Phi_0 =0$,
with $\mbf J = (c/4\pi) \nabla\times \mbf B$.
Taking into account the continuity equation
$\partial B/\partial t + \nabla\cdot(B \mbf v) = 0$,
the dynamics of the magnetic flux in a superconductor is
described by the well-known nonlinear diffusion equation%
\cite{vinokur_exact_1991,abulafia_local_1995,ProzorovPRB58-98}
\begin{equation}\label{eq:NonlinBdiff}
\frac{\partial B}{\partial t} =
\frac{\Phi_0}{4\pi\eta}\; \nabla\cdot (|B|\; \nabla B).
\end{equation}

\begin{figure}[b]
\includegraphics[scale=0.68]{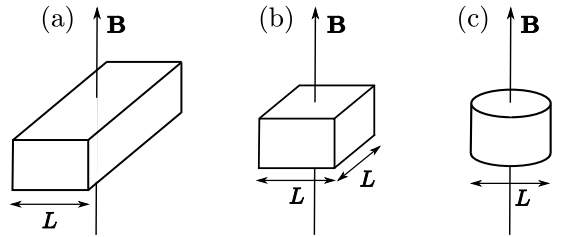}
\caption{\label{fig:geometries.eps}
    Vortex dynamics is studied for three sample geometries:
    (a) a slab, (b) a square-shaped plate, and (c) a disk.
}
\end{figure}

We have solved Eq.~\eqref{eq:NonlinBdiff} for three sample
geometries: a slab, a square-shaped plate, and a disk
(see Fig.~\ref{fig:geometries.eps}).
We assume the sample thickness along the field is sufficiently large
and neglect stray fields on the top and bottom of the sample.
Magnetic induction $\mbf B (\mbf r, t)$ is
directed along the sample symmetry axis $z$ and
satisfies the following initial and boundary conditions: (i)
$B$ is uniform within the sample at $t=0$,
$B(x,y; t=0) = B_0$, and (ii) $B$ vanishes at the
sample edges for $t>0$.

For a long superconducting slab of width $L$,
Eq. \eqref{eq:NonlinBdiff} reads
\begin{equation}\label{eq:NonlinBdiffStrip}
\frac{\partial B}{\partial t} =
\frac{\Phi_0}{4\pi\eta}\;
\frac{\partial}{\partial x}
\Big( |B|\; \frac{\partial B}{\partial x} \Big),
\end{equation}
where 
$B(x,t)= 0$ at $x=\pm L/2$ for $t>0$. We can seek the
solution in the form
\begin{equation}\label{eq:Bxt-series}
B(x,t) = \sum_{k=1}^\infty B_k(t) \sin[k\pi(x/L+1/2)],
\end{equation}
with functions $B_k(t)$ to be determined from
Eq.~\eqref{eq:NonlinBdiffStrip} and $B(x,t=0)=B_0$.
This gives the following set of differential equations:
\begin{equation}\label{eq:bn-diffEq}
\frac{d B_k(t)}{dt}
= \frac{1}{B_0\tau_0}\sum_{i,j=1}^\infty
B_i(t)\, F_k(i,j)\, B_j(t),
\end{equation}
with the initial conditions $B_k(0) = (2B_0/\pi k)\, [1-(-1)^k]$
($k=1,2,\ldots$). Here, the coefficients $F_k(i,j)$
are given by
\begin{equation}
F_k(i,j)
=
\frac{k\pi}{4}
\left(
    \frac{(i-j)^2}{(i-j)^2-k^2} - \frac{(i+j)^2}{(i+j)^2-k^2}
\right)
\end{equation}
for $|i\pm j| \ne k$ and $i+j+k$ odd, and $F_k(i,j)=0$ otherwise.
The characteristic time constant is
\begin{equation}\label{eq:tau0}
\tau_0 = \frac{\pi L^2 \eta}{\Phi_0 B_0}.
\end{equation}
Equations \eqref{eq:bn-diffEq} are solved by truncating the system
at sufficiently large $k$ ($k\sim 40$).
%
\begin{figure}[t]
\includegraphics[scale=0.59]{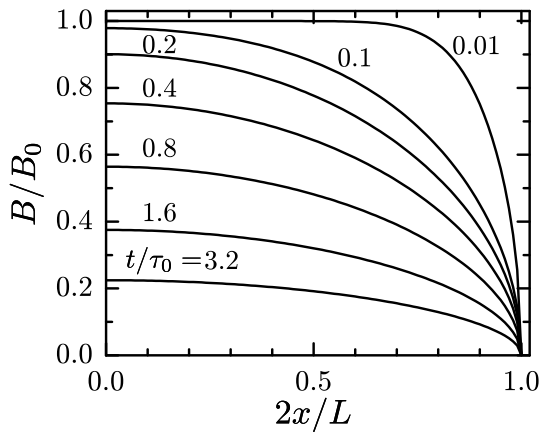}
\caption{\label{fig:BofXStrip.eps}
    Magnetic induction $B(x)$ across one half of a superconducting slab
    of width $L$ is shown for times $t/\tau_0 =0.01$, $0.1$, $0.2$,\ldots,
    $3.2$ (top to bottom).
}
\end{figure}
%
The induction $B(x,t)$ for the slab is shown in
Fig.~\ref{fig:BofXStrip.eps} at various times $t/\tau_0 = 0.01$,
$0.1$, $0.2$, ..., $3.2$. We observe that the flux flow near the
sample edges in the initial time interval is very fast, reaching
the center of the slab ($x=0$) at time $t \sim 0.1\, \tau_0$ after
switching off the field. This regime is followed by a slower
flux flow taking place in the bulk of the sample.

The spatial dependence of the magnetic induction
is in accordance with the previous results for the flux flow regime
with constant activation energy.\cite{ProzorovPRB58-98}
In the presence of flux creep, which may take place in the center of
the slab for $t\ll 0.1\tau_0$, or at large times $t\gg \tau_0$
when remanent magnetization is small, a phenomenological
model of current and field-dependent activation energy
should be used.%
\cite{feigelman_theory_1989,vinokur_exact_1991,abulafia_local_1995,ProzorovPRB58-98,gurevich_time_1993}
Note that the obtained $B(x,t)$ shown in Fig.~\ref{fig:BofXStrip.eps}
is qualitatively different from the solution of Eq.~\eqref{eq:NonlinBdiffStrip}
when the field is switched on at $t=0$.
In that case the magnetic field enters the sample in the form of a
flux front propagating from the edges.%
\cite{vinokur_exact_1991,ProzorovPRB58-98,%
LandauVol6FluidMechanics,*bass_nonlinear_1998}
Magnetic induction in the vicinity of the front is a linear function of the
coordinate, $B(x,t) = (4\pi\eta/\Phi_0)\, v_f |x-x_f|$, with
$x_f(t)$ and $v_f$ being the position and the velocity of the front.
In our case, the field is switched off
at $t=0$ and the flux escapes the sample with no front in $B(x,t)$
formed even at $t\ll \tau_0$. Indeed, at a sufficiently large
distance $u$ from the edge, Eq.~\eqref{eq:NonlinBdiffStrip}
can be linearized with respect to
$\delta B =B_0 - B$, which gives the exponential decay
$\delta B(u,t) \propto (u/2\kappa\sqrt{t})^{-1}\,
e^{-u^2/4\kappa^2 t}$ ($\kappa = \sqrt{\Phi_0 B_0/4\pi\eta}$)
characteristic of the linear diffusion.

\begin{figure}[t]
\includegraphics[scale=0.80]{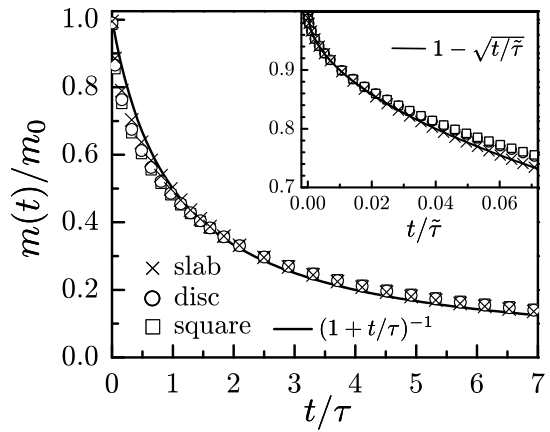}
\caption{\label{fig:BavgFIT1.eps}
    Numerical solutions for magnetic moment relaxation for the slab
    ($\times$), disk ({\large$\circ$}), and square
    ({\scriptsize$\Box$}) geometries, compared to the analytic
    approximation, Eq.~\eqref{eq:Mapprox} (solid line).
    Inset: A better fit for the initial time interval $t\ll \tau$
    to the analytic expression, Eq.~\eqref{eq:m_edges}.
    In this case, only magnetization at the edges is
    affected by the flux flow.
}
\end{figure}

In the following we study the dynamics of the average magnetic
induction $\bar B(t) = A^{-1} \int dxdy\; B(x,y,t)$
($A$ is the sample area) which is proportional to
the magnetic moment $m(t)$ that can be measured.
There are two regimes
of the flux dynamics in the system. At very short times
$t\ll \tau_0$ after switching off the field, the flux flow is localized
near the edges and is unaffected by the sample size.
In this case, the solution for a half-infinite
superconductor is a good approximation,
$B(u,t) = B_0 f(u/\kappa\sqrt{t})$.\cite{bryksin_nonlinear_1993}
Here, $f$ is a dimensionless function to be determined from
Eq.~\eqref{eq:NonlinBdiffStrip} for the half-infinite superconductor
with the boundary conditions $f(0)=0$ and $f(\infty)=1$.
Using the above expression for $B(u,t)$ and taking into account
that it deviates significantly from $B_0$ in the vicinity of the edges,
we find for the average induction $1-\bar B(t)/B_0 \propto
(P/A)\kappa\sqrt{t}$, where $P$ is the
perimeter of the sample. This gives the magnetic moment
relaxation
\begin{equation}\label{eq:m_edges}
m(t) = m_0 \left( 1 - \sqrt{t/\tilde \tau}\right)\,,\qquad t \ll \tilde\tau\,,
\end{equation}
with the time constant
\begin{equation}
\tilde \tau = 9\pi (A/P)^2 \eta/ \Phi_0 B_0,
\end{equation}
where the numerical prefactor characterizes the spatial spread
of $B$ away from the edges. Comparison with the numerical solution
for $m(t)$ is shown in the inset of Fig.~\ref{fig:BavgFIT1.eps} for different
sample geometries. We find that Eq.~\eqref{eq:m_edges} is a good
approximation of the exact $m(t)$ in the short initial time
interval $t/\tilde \tau\lesssim 0.1$ before the flux flow
reaches the center of the sample. The flux flow in this time
interval is very fast, leading to a $30\%$ reduction of the
overall magnetic moment. 

\begin{figure}[t]
\includegraphics[scale=0.78]{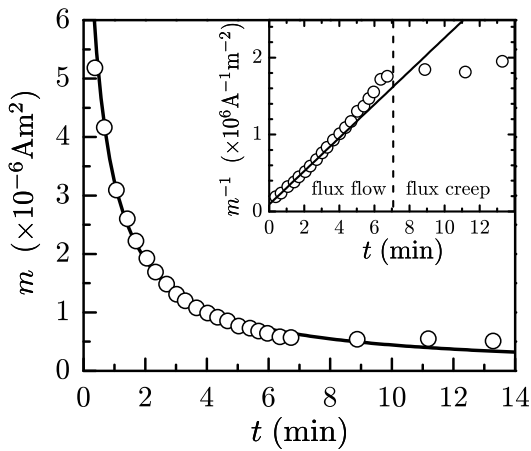}
\caption{\label{mvst-Moshchalkov.eps}
    Experimental data\cite{moshchalkov_early_1989}
    ({\large$\circ$}) for the magnetic moment $m(t)$
    in BSCO single crystal fitted to
    Eq.~\eqref{eq:Mapprox} with $m_0 = 1.1\times 10^{-5}\rm{Am}^2$ and
    $\tau = 0.43\,$min (solid curve). This corresponds to
    $\eta =0.5\, \rm{Ns}/\rm m^2$.
    Inset: The inverse magnetic moment as a function of time.
    The crossover between flux flow and
    flux creep regimes is seen as a dramatic change of the slope
    at $t \approx 7\,$min (dashed line).
    The sample is a slab $0.14\times 1.37\times 2.06\,$mm in size,
    the initial magnetic induction $B_0 = 35\,$mT, and $T=77$K.
}
\end{figure}
At times $t\gtrsim \tau_0$ the flux flow extends through the whole
sample, giving rise to the magnetization relaxation which depends
on geometry. For the superconducting slab, the
first-order approximation of Eqs.~\eqref{eq:bn-diffEq} for $k=1$ reads
$\bar B^{(1)}(t) = (8 B_0/\pi^2) [1+t/(0.75 \tau_0)]^{-1}$.
Truncating Eqs.~\eqref{eq:bn-diffEq} at $k\sim 40$,
a practically exact solution is obtained.
This solution can be approximated by a simple formula,
$\bar B(t) = B_0 [1+t/(0.62 \tau_0)]^{-1}$,
which is very close to the exact one for $t\gtrsim \tau_0$.
This suggests that the exact solution for the magnetic moment $m(t)$
can be accurately approximated by
\begin{equation}\label{eq:Mapprox}
m(t) = \frac{m_0}{1+t/\tau}, \qquad \tau = \alpha\tau_0,
\end{equation}
where $\alpha$ is a number which depends on geometry.
Fitting the exact numerical solution for $m(t)$ to Eq.~\eqref{eq:Mapprox}
we find $\alpha=0.620$, $0.244$, and $0.226$ for the slab, square,
and disk geometries, respectively (Fig.~\ref{fig:BavgFIT1.eps}).
The fitting ensures the smallest absolute error between
exact and fitted $m(t)$ for $1 < t/\tau < 3$.
As expected, the decay of $m(t)$ is slower (that is, geometric
factor $\alpha$ is larger) for the slab than for the disk,
other parameters being equal.


In what follows, we analyze available experimental data on $m(t)$
and extract the characteristic time constant as well as the effective drag $\eta$.
Relaxation of the magnetic moment
in BSCO single crystals is studied in Ref.~\onlinecite{moshchalkov_early_1989}.
Experimental data
are shown in Fig.~\ref{mvst-Moshchalkov.eps} (open circles)
fitted to Eq.~\eqref{eq:Mapprox} (solid curve)
with $m_0 = 1.1\times 10^{-5} \rm A \rm m^2$ and
$\tau = 0.43\; \rm{min}$.
The fitting is performed for
the initial time interval before logarithmic in time, thermally activated
flux creep sets in.
The linear time dependence of the inverse magnetic moment
is shown in the inset of Fig.~\ref{mvst-Moshchalkov.eps};
the crossover between flux flow and flux creep regimes is seen as a
dramatic change of the slope at $t/ \tau \approx 16$.

Let us now extract $\eta$.
The dimensions of the sample used in the experiment
are $0.14\times 1.37\times 2.06 \; \rm{mm}$,
which gives $B_0=4\pi m_0/V = 35\,$mT, where $V$ is the volume.
Taking $\alpha = 0.620$ for the slab of the width
$L=0.14\; \rm{mm}$, we obtain $\eta = 0.5\; \rm{Ns/m}^2$.
This value for the effective vortex viscosity exceeds
by six orders of magnitude the Bardeen-Stephen
drag coefficient $\eta_0\sim 10^{-7}\, \rm {Ns}/\rm{m}^2$
measured in BSCO.\cite{GolosovskyDavidovSUPERCONDSCITECHNOL9-96}
Huge enhancement of the drag indicates a strong influence of
the pinning on the vortex diffusion. Despite the
complexity of the vortex motion in the presence of pinning
and thermal agitation, the magnetization
follows a simple algebraic time dependence, Eq.~\eqref{eq:Mapprox}.

\begin{figure}[t]
\includegraphics[scale=0.60]{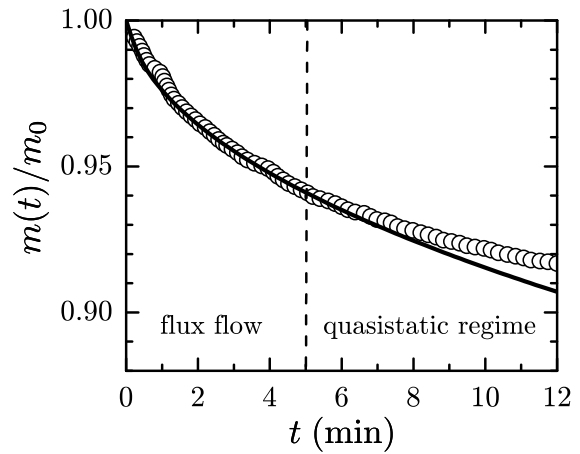}
\caption{\label{BishopFitExact.eps}
    Experimental data\cite{pardo_observation_1998} ({\large$\circ$})
    for magnetic moment relaxation in $\rm{NbSe}_2$ monocrystal
    fitted to Eq.~\eqref{eq:m_edges} (solid curve) with
    $\tilde \tau = 1.13\times 10^3$min,
    corresponding to $\eta = 11\, \rm {Ns}/\rm m^2$.
    The dashed line indicates a crossover between the flux
    flow regime, Eq.~\eqref{eq:NonlinBdiff},
    and the slow quasistatic motion before the flux creep.
    The sample has a square geometry $0.5\times 0.5\times 0.2\, \rm{mm}$,
    the initial magnetic induction $B_0=3.3\,\rm{mT}$, and $T=4.2$K.
}
\end{figure}

Vortex dynamics has been studied in $\rm{NbSe}_2$
using the decoration technique for visualization of flowing vortex
lattices.\cite{pardo_observation_1998}
Magnetization measurements have been performed using the SQUID
(superconducting quantum interference device) magnetometry.
A crossover has been observed as a function of increasing flux density
from a layered (smectic) flowing flux lattice in the disorder-dominated
low-field limit to a more ordered (Bragg glass) lattice structure in the
interaction-dominated high-field case.
The observed time dependence of
magnetization relaxation in the high-field limit ($B_0 = 3.3\,$mT)
is shown in Fig.~\ref{BishopFitExact.eps}. The regimes indicated
in Fig.~\ref{BishopFitExact.eps} correspond to
the flux flow and to the quasistatic vortex motion.
The solid curve in Fig.~\ref{BishopFitExact.eps} is the fit of $m(t)$
to Eq.~\eqref{eq:m_edges} for the $\rm{NbSe}_2$ sample
$0.5\times 0.5\times 0.2\,$mm in size, which gives the
relaxation time $\tilde \tau = 1.13\times 10^3\,$min and the
viscous drag coefficient $\eta=11\,\rm{Ns}/\rm{m}^2$.
We observe that the simple hydrodynamic model with an effective
viscous drag force fits the data in the initial
stages of magnetization relaxation where the vortex density is large
and the flux flow takes place.
The flux flow is localized near the edges, as corroborated
experimentally by a small reduction of the magnetic moment of the
sample over the measurement time and, more directly, by observing
the static vortex structure in the center of the
sample.\cite{pardo_observation_1998} Large effective $\eta$
is clearly due to hopping caused by successive pinning and
thermally-assisted 
depinning of vortices, as evidenced 
by studying single-vortex dynamics in pristine $\rm{NbSe}_2$
monocrystals by scanning tunneling
microscopy.\cite{troyanovski_collective_1999}

\begin{figure}[t]
\includegraphics[scale=0.82]{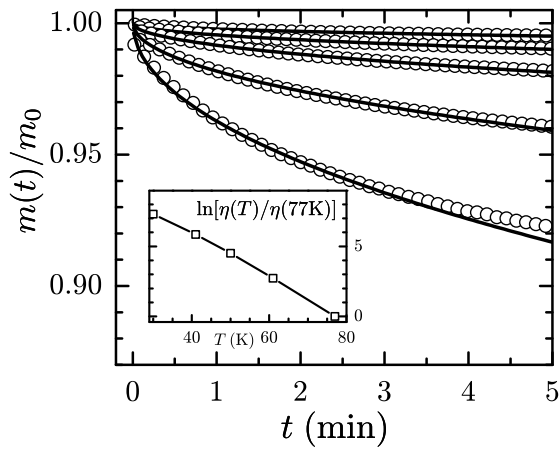}
\caption{\label{DengTsuzukiHara.eps}
    Experimental data\cite{deng_relaxation_2012} ({\large $\circ$})
    for magnetic moment relaxation in YBCO polycrystal
    at temperatures $T = 30$, $41$, $50$, $61$, and $77\,$K
    (top to bottom), fitted to Eq.~\eqref{eq:m_edges}
    (solid curves) with
    $\tilde\tau = 2.1\times 10^5$, $5.0\times 10^4$, $1.4\times 10^4$,
    $3.0\times 10^3$, and $7.4\times 10^2$ min, respectively.
    This corresponds to the drag coefficients
    $\eta = 284$, $66$, $17$, $2.93$, and $0.19$ $\rm{Ns/m}^2$.
    Inset: Logarithm of the drag coefficient, normalized to
    $\eta(77\rm K) = 0.19\, \rm{Ns/m}^2$, as a function of
    the temperature.
    The sample has a rectangular geometry of $66\times 34\times 15\,$mm.
    The initial magnetic induction is $B_0 = 3.95$, $3.77$, $3.48$, $2.80$,
    $0.735\,$T, respectively.
}
\end{figure}

\begin{figure}[t]
\includegraphics[scale=0.82]{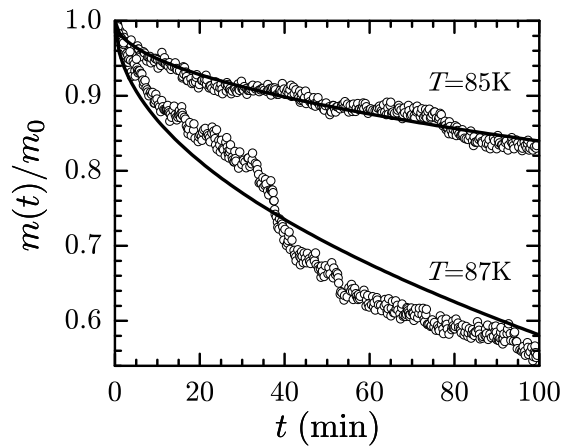}
\caption{\label{Monarkha.eps}
    Experimental data\cite{monarkha_violation_2012} ({\large $\circ$})
    for magnetic moment relaxation in YBCO monocrystal
    at temperatures of $85$ and $87\,$K
    (top to bottom), fitted to Eq.~\eqref{eq:m_edges}
    with $\tilde \tau = 3.8\times 10^3\,$min and
    $\tilde\tau = 570\,$min, respectively (solid curves).
    The corresponding drag coefficients are
    $\eta = 0.27$ and $0.04$ $\rm{Ns/m}^2$.
    The sample has a square geometry of $1\times 1\times 0.02\,$mm,
    and the initial magnetic induction is $B_0 = 0.1\, \rm{mT}$.
}
\end{figure}

Magnetic moment relaxations in YBCO
polycrystal\cite{deng_relaxation_2012} and
monocrystal\cite{monarkha_violation_2012}
are shown in Figs.~\ref{DengTsuzukiHara.eps}
and~\ref{Monarkha.eps}.
The relaxation in the polycrystalline YBCO (rectangular geometry,
$66\times 34\times 15$mm in size) is studied at
$30$, $41$, $50$, $61$, and $77$K.
The initial stage of magnetization relaxation can be fitted by
Eq.~\eqref{eq:m_edges} describing the flux flow in the vicinity of
the edges (Fig.~\ref{DengTsuzukiHara.eps}, solid curves). This is
in agreement with the observed small reduction of the overall
magnetic moment during the measurement.
The obtained effective viscosity strongly depends
on temperature, ranging between $\eta\sim 100$
and $0.1\,\rm{Ns}/\rm{m}^2$ as the temperature is increased
from $30$K to $77$K, see inset of Fig.~\ref{DengTsuzukiHara.eps}.
The extracted value $\eta(77\rm{K}) = 0.19\,\rm{Ns}/\rm{m}^2$
is consistent with the value $\eta=0.12\,\rm{Ns}/\rm{m}^2$ measured
independently at the same temperature by studying the spatiotemporal
change of the magnetization profile in a bulk YBCO sample in
the flux-flow regime.\cite{bondarenko_study_2006}
Taking $\eta \propto e^{U/kT}$ and neglecting the temperature dependence
of the effective activation energy $U$ as well as of the prefactor, we find
$U\approx 360\,\rm{K}$ in accordance with the previous
results.\cite{abulafia_local_1995}

Magnetic moment relaxation in small YBCO
monocrystal ($1\times 1\times 0.02\, \rm{mm}$) is shown in
Fig.~\ref{Monarkha.eps}.\cite{monarkha_violation_2012}
The data can be fitted with the effective
viscous drag coefficients $\eta=0.27$ and $0.04\,\rm{Ns}/\rm{m}^2$
at temperatures of $85$K and $87$K, respectively. The decrease
of $\eta$ in such a narrow temperature range may be due to the
proximity of the critical temperature ($T_c\approx 88$K) where
fluctuations are pronounced. In addition, the sharp change in the
relaxation rate observed at $87$K and $t\approx 30\,$min
suggests that the flux flow is inhomogeneous and made of large
domains which, upon depinning, abruptly increase the magnetic
moment relaxation rate.

In conclusion, we have studied vortex dynamics in type-II
superconductors in the initial time interval before the flux creep
sets in. We have used a simple phenomenological (hydrodynamic) model
of nonlinear diffusion of massless vortices where pinning of the
flux lines by material inhomogeneities, interaction with other
vortices and the surface, and the Bardeen-Stephen dissipation
in the vortex core are described by
an effective viscous drag coefficient, $\eta$.
After switching off the external magnetic field,
the vortex dynamics exhibits two distinct regimes
before the creep sets in with logarithmic in time decay of
remanent magnetization.
In the beginning, the flux flow is localized near the edges
and is independent of the sample size. At later times,
this regime is followed by a slower flux flow
involving the bulk of the sample.
We find that magnetic moment
relaxation in these regimes can be accurately approximated by
$m(t)=m_0(1-\sqrt{t/\tilde \tau})$ for $t\ll \tilde \tau$ and
$m(t)=m_0(1+t/\tau)^{-1}$ for $t\gtrsim \tau$, where geometry-dependent
$\tilde\tau$ and $\tau$ are proportional to $\eta$.

We have analyzed available experimental data
on early stages of magnetization relaxation after the magnetic field
is instantaneously removed. We obtained
quantitative agreement for both conventional and high-temperature
superconductors, albeit with $\eta$ exceeding the Bardeen-Stephen
value $\eta_0$ by many orders of magnitude. Huge enhancement of
$\eta$ with respect to $\eta_0$,
as well as its exponential temperature dependence,
indicates a strong influence of pinning and thermally assisted
depinning of vortices on flux diffusion.
We argue that early stages of magnetization relaxation
can be modeled as a flux flow with an effective
drag coefficient.
This allows for a simple experimental
determination of the bulk vortex viscosity, which cannot be
accessed by the surface impedance measurements.


This research was supported by the Serbian Ministry of Science,
Project No.~171027. Work by V.K. at the Ames Laboratory is supported
by the U.S. Department of Energy, Office of  Basic Energy Sciences,
Division of Materials Sciences and Engineering under Contract No.
DE-AC02-07CH11358. M.V. acknowledges support by the DFG through SFB
767 and the hospitality of the Quantum Transport Group,
Universit{\" a}t Konstanz, Germany, where part of this work was done.


\bibliography{References}

\begin{thebibliography}{31}%
\makeatletter
\providecommand \@ifxundefined [1]{%
 \@ifx{#1\undefined}
}%
\providecommand \@ifnum [1]{%
 \ifnum #1\expandafter \@firstoftwo
 \else \expandafter \@secondoftwo
 \fi
}%
\providecommand \@ifx [1]{%
 \ifx #1\expandafter \@firstoftwo
 \else \expandafter \@secondoftwo
 \fi
}%
\providecommand \natexlab [1]{#1}%
\providecommand \enquote  [1]{``#1''}%
\providecommand \bibnamefont  [1]{#1}%
\providecommand \bibfnamefont [1]{#1}%
\providecommand \citenamefont [1]{#1}%
\providecommand \href@noop [0]{\@secondoftwo}%
\providecommand \href [0]{\begingroup \@sanitize@url \@href}%
\providecommand \@href[1]{\@@startlink{#1}\@@href}%
\providecommand \@@href[1]{\endgroup#1\@@endlink}%
\providecommand \@sanitize@url [0]{\catcode `\\12\catcode `\$12\catcode
  `\&12\catcode `\#12\catcode `\^12\catcode `\_12\catcode `\%12\relax}%
\providecommand \@@startlink[1]{}%
\providecommand \@@endlink[0]{}%
\providecommand \url  [0]{\begingroup\@sanitize@url \@url }%
\providecommand \@url [1]{\endgroup\@href {#1}{\urlprefix }}%
\providecommand \urlprefix  [0]{URL }%
\providecommand \Eprint [0]{\href }%
\providecommand \doibase [0]{http://dx.doi.org/}%
\providecommand \selectlanguage [0]{\@gobble}%
\providecommand \bibinfo  [0]{\@secondoftwo}%
\providecommand \bibfield  [0]{\@secondoftwo}%
\providecommand \translation [1]{[#1]}%
\providecommand \BibitemOpen [0]{}%
\providecommand \bibitemStop [0]{}%
\providecommand \bibitemNoStop [0]{.\EOS\space}%
\providecommand \EOS [0]{\spacefactor3000\relax}%
\providecommand \BibitemShut  [1]{\csname bibitem#1\endcsname}%
\let\auto@bib@innerbib\@empty
\bibitem [{\citenamefont {Crabtree}\ and\ \citenamefont
  {Nelson}(1997)}]{CrabtreeNelsonPHYSTODAY50-97}%
  \BibitemOpen
  \bibfield  {author} {\bibinfo {author} {\bibfnamefont {G.~W.}\ \bibnamefont
  {Crabtree}}\ and\ \bibinfo {author} {\bibfnamefont {D.~R.}\ \bibnamefont
  {Nelson}},\ }\href@noop {} {\bibfield  {journal} {\bibinfo  {journal} {Phys.
  Today}\ }\textbf {\bibinfo {volume} {50}},\ \bibinfo {pages} {38} (\bibinfo
  {year} {1997})}\BibitemShut {NoStop}%
\bibitem [{\citenamefont {Brandt}(1995)}]{BrandtRPROGPHYS58-95}%
  \BibitemOpen
  \bibfield  {author} {\bibinfo {author} {\bibfnamefont {E.~H.}\ \bibnamefont
  {Brandt}},\ }\href@noop {} {\bibfield  {journal} {\bibinfo  {journal} {Rep.
  Prog. Phys.}\ }\textbf {\bibinfo {volume} {58}},\ \bibinfo {pages} {1465}
  (\bibinfo {year} {1995})}\BibitemShut {NoStop}%
\bibitem [{\citenamefont {Anderson}\ and\ \citenamefont
  {Kim}(1964)}]{AndersonKimRMP36-64}%
  \BibitemOpen
  \bibfield  {author} {\bibinfo {author} {\bibfnamefont {P.~W.}\ \bibnamefont
  {Anderson}}\ and\ \bibinfo {author} {\bibfnamefont {Y.~B.}\ \bibnamefont
  {Kim}},\ }\href@noop {} {\bibfield  {journal} {\bibinfo  {journal} {Rev. Mod.
  Phys.}\ }\textbf {\bibinfo {volume} {36}},\ \bibinfo {pages} {39} (\bibinfo
  {year} {1964})}\BibitemShut {NoStop}%
\bibitem [{\citenamefont {Bardeen}\ and\ \citenamefont
  {Stephen}(1965)}]{bardeen_theory_1965}%
  \BibitemOpen
  \bibfield  {author} {\bibinfo {author} {\bibfnamefont {J.}~\bibnamefont
  {Bardeen}}\ and\ \bibinfo {author} {\bibfnamefont {M.~J.}\ \bibnamefont
  {Stephen}},\ }\href@noop {} {\bibfield  {journal} {\bibinfo  {journal} {Phys.
  Rev.}\ }\textbf {\bibinfo {volume} {140}},\ \bibinfo {pages} {A1197}
  (\bibinfo {year} {1965})}\BibitemShut {NoStop}%
\bibitem [{\citenamefont {Beasley}\ \emph {et~al.}(1969)\citenamefont
  {Beasley}, \citenamefont {Labusch},\ and\ \citenamefont
  {Webb}}]{beasley_flux_1969}%
  \BibitemOpen
  \bibfield  {author} {\bibinfo {author} {\bibfnamefont {M.~R.}\ \bibnamefont
  {Beasley}}, \bibinfo {author} {\bibfnamefont {R.}~\bibnamefont {Labusch}}, \
  and\ \bibinfo {author} {\bibfnamefont {W.~W.}\ \bibnamefont {Webb}},\
  }\href@noop {} {\bibfield  {journal} {\bibinfo  {journal} {Phys. Rev.}\
  }\textbf {\bibinfo {volume} {181}},\ \bibinfo {pages} {682} (\bibinfo {year}
  {1969})}\BibitemShut {NoStop}%
\bibitem [{\citenamefont {Tinkham}(1996)}]{TinkhamBook}%
  \BibitemOpen
  \bibfield  {author} {\bibinfo {author} {\bibfnamefont {M.}~\bibnamefont
  {Tinkham}},\ }\href@noop {} {\emph {\bibinfo {title} {Introduction to
  Superconductivity}}}\ (\bibinfo  {publisher} {McGraw-Hill},\ \bibinfo
  {address} {New York},\ \bibinfo {year} {1996})\BibitemShut {NoStop}%
\bibitem [{\citenamefont {Blatter}\ \emph {et~al.}(1994)\citenamefont
  {Blatter}, \citenamefont {Feigel'man}, \citenamefont {Geshkenbein},
  \citenamefont {Larkin},\ and\ \citenamefont
  {Vinokur}}]{blatter_vortices_1994}%
  \BibitemOpen
  \bibfield  {author} {\bibinfo {author} {\bibfnamefont {G.}~\bibnamefont
  {Blatter}}, \bibinfo {author} {\bibfnamefont {M.~V.}\ \bibnamefont
  {Feigel'man}}, \bibinfo {author} {\bibfnamefont {V.~B.}\ \bibnamefont
  {Geshkenbein}}, \bibinfo {author} {\bibfnamefont {A.~I.}\ \bibnamefont
  {Larkin}}, \ and\ \bibinfo {author} {\bibfnamefont {V.~M.}\ \bibnamefont
  {Vinokur}},\ }\href@noop {} {\bibfield  {journal} {\bibinfo  {journal} {Rev.
  Mod. Phys.}\ }\textbf {\bibinfo {volume} {66}},\ \bibinfo {pages} {1125}
  (\bibinfo {year} {1994})}\BibitemShut {NoStop}%
\bibitem [{\citenamefont {Yeshurun}\ \emph {et~al.}(1996)\citenamefont
  {Yeshurun}, \citenamefont {Malozemoff},\ and\ \citenamefont
  {Shaulov}}]{yeshurun_magnetic_1996}%
  \BibitemOpen
  \bibfield  {author} {\bibinfo {author} {\bibfnamefont {Y.}~\bibnamefont
  {Yeshurun}}, \bibinfo {author} {\bibfnamefont {A.~P.}\ \bibnamefont
  {Malozemoff}}, \ and\ \bibinfo {author} {\bibfnamefont {A.}~\bibnamefont
  {Shaulov}},\ }\href@noop {} {\bibfield  {journal} {\bibinfo  {journal} {Rev.
  Mod. Phys.}\ }\textbf {\bibinfo {volume} {68}},\ \bibinfo {pages} {911}
  (\bibinfo {year} {1996})}\BibitemShut {NoStop}%
\bibitem [{\citenamefont {Marchetti}\ and\ \citenamefont
  {Nelson}(1990)}]{marchetti_hydrodynamics_1990}%
  \BibitemOpen
  \bibfield  {author} {\bibinfo {author} {\bibfnamefont {M.~C.}\ \bibnamefont
  {Marchetti}}\ and\ \bibinfo {author} {\bibfnamefont {D.~R.}\ \bibnamefont
  {Nelson}},\ }\href@noop {} {\bibfield  {journal} {\bibinfo  {journal} {Phys.
  Rev. B}\ }\textbf {\bibinfo {volume} {42}},\ \bibinfo {pages} {9938}
  (\bibinfo {year} {1990})}\BibitemShut {NoStop}%
\bibitem [{\citenamefont {Okada}\ \emph {et~al.}(2012)\citenamefont {Okada},
  \citenamefont {Takahashi}, \citenamefont {Imai}, \citenamefont {Kitagawa},
  \citenamefont {Matsubayashi}, \citenamefont {Uwatoko},\ and\ \citenamefont
  {Maeda}}]{okada_PRB86-12}%
  \BibitemOpen
  \bibfield  {author} {\bibinfo {author} {\bibfnamefont {T.}~\bibnamefont
  {Okada}}, \bibinfo {author} {\bibfnamefont {H.}~\bibnamefont {Takahashi}},
  \bibinfo {author} {\bibfnamefont {Y.}~\bibnamefont {Imai}}, \bibinfo {author}
  {\bibfnamefont {K.}~\bibnamefont {Kitagawa}}, \bibinfo {author}
  {\bibfnamefont {K.}~\bibnamefont {Matsubayashi}}, \bibinfo {author}
  {\bibfnamefont {Y.}~\bibnamefont {Uwatoko}}, \ and\ \bibinfo {author}
  {\bibfnamefont {A.}~\bibnamefont {Maeda}},\ }\href@noop {} {\bibfield
  {journal} {\bibinfo  {journal} {Phys. Rev. B}\ }\textbf {\bibinfo {volume}
  {86}},\ \bibinfo {pages} {064516} (\bibinfo {year} {2012})}\BibitemShut
  {NoStop}%
\bibitem [{\citenamefont {Raes}\ \emph {et~al.}(2012)\citenamefont {Raes},
  \citenamefont {Van~de Vondel}, \citenamefont {Silhanek}, \citenamefont
  {de~Souza~Silva}, \citenamefont {Gutierrez}, \citenamefont {Kramer},\ and\
  \citenamefont {Moshchalkov}}]{RaesMoshchalkovPRB86-12}%
  \BibitemOpen
  \bibfield  {author} {\bibinfo {author} {\bibfnamefont {B.}~\bibnamefont
  {Raes}}, \bibinfo {author} {\bibfnamefont {J.}~\bibnamefont {Van~de Vondel}},
  \bibinfo {author} {\bibfnamefont {A.~V.}\ \bibnamefont {Silhanek}}, \bibinfo
  {author} {\bibfnamefont {C.~C.}\ \bibnamefont {de~Souza~Silva}}, \bibinfo
  {author} {\bibfnamefont {J.}~\bibnamefont {Gutierrez}}, \bibinfo {author}
  {\bibfnamefont {R.~B.~G.}\ \bibnamefont {Kramer}}, \ and\ \bibinfo {author}
  {\bibfnamefont {V.~V.}\ \bibnamefont {Moshchalkov}},\ }\href@noop {}
  {\bibfield  {journal} {\bibinfo  {journal} {Phys. Rev. B}\ }\textbf {\bibinfo
  {volume} {86}},\ \bibinfo {pages} {064522} (\bibinfo {year}
  {2012})}\BibitemShut {NoStop}%
\bibitem [{\citenamefont {Lin}\ \emph {et~al.}(2012)\citenamefont {Lin},
  \citenamefont {Bulaevskii},\ and\ \citenamefont {Batista}}]{lin_vortex_2012}%
  \BibitemOpen
  \bibfield  {author} {\bibinfo {author} {\bibfnamefont {S.-Z.}\ \bibnamefont
  {Lin}}, \bibinfo {author} {\bibfnamefont {L.~N.}\ \bibnamefont {Bulaevskii}},
  \ and\ \bibinfo {author} {\bibfnamefont {C.~D.}\ \bibnamefont {Batista}},\
  }\href@noop {} {\bibfield  {journal} {\bibinfo  {journal} {Phys. Rev. B}\
  }\textbf {\bibinfo {volume} {86}},\ \bibinfo {pages} {180506} (\bibinfo
  {year} {2012})}\BibitemShut {NoStop}%
\bibitem [{\citenamefont {Kunchur}\ \emph {et~al.}(1993)\citenamefont
  {Kunchur}, \citenamefont {Christen},\ and\ \citenamefont
  {Phillips}}]{kunchur-PRL70-93}%
  \BibitemOpen
  \bibfield  {author} {\bibinfo {author} {\bibfnamefont {M.~N.}\ \bibnamefont
  {Kunchur}}, \bibinfo {author} {\bibfnamefont {D.~K.}\ \bibnamefont
  {Christen}}, \ and\ \bibinfo {author} {\bibfnamefont {J.~M.}\ \bibnamefont
  {Phillips}},\ }\href@noop {} {\bibfield  {journal} {\bibinfo  {journal}
  {Phys. Rev. Lett.}\ }\textbf {\bibinfo {volume} {70}},\ \bibinfo {pages}
  {998} (\bibinfo {year} {1993})}\BibitemShut {NoStop}%
\bibitem [{\citenamefont {Moshchalkov}\ \emph {et~al.}(1989)\citenamefont
  {Moshchalkov}, \citenamefont {Zhukov}, \citenamefont {Kuznetsov},
  \citenamefont {Metlushko},\ and\ \citenamefont
  {Leonyuk}}]{moshchalkov_early_1989}%
  \BibitemOpen
  \bibfield  {author} {\bibinfo {author} {\bibfnamefont {V.}~\bibnamefont
  {Moshchalkov}}, \bibinfo {author} {\bibfnamefont {A.}~\bibnamefont {Zhukov}},
  \bibinfo {author} {\bibfnamefont {V.}~\bibnamefont {Kuznetsov}}, \bibinfo
  {author} {\bibfnamefont {V.}~\bibnamefont {Metlushko}}, \ and\ \bibinfo
  {author} {\bibfnamefont {L.}~\bibnamefont {Leonyuk}},\ }\href@noop {}
  {\bibfield  {journal} {\bibinfo  {journal} {{JETP} Lett.}\ }\textbf {\bibinfo
  {volume} {50}},\ \bibinfo {pages} {91} (\bibinfo {year} {1989})}\BibitemShut
  {NoStop}%
\bibitem [{\citenamefont {Moshchalkov}\ \emph {et~al.}(1991)\citenamefont
  {Moshchalkov}, \citenamefont {Zhukov}, \citenamefont {Kuznetsov},
  \citenamefont {Metlushko},\ and\ \citenamefont
  {Leonyuk}}]{MoshchalkovPHYSICAB169-91}%
  \BibitemOpen
  \bibfield  {author} {\bibinfo {author} {\bibfnamefont {V.}~\bibnamefont
  {Moshchalkov}}, \bibinfo {author} {\bibfnamefont {A.}~\bibnamefont {Zhukov}},
  \bibinfo {author} {\bibfnamefont {V.}~\bibnamefont {Kuznetsov}}, \bibinfo
  {author} {\bibfnamefont {V.}~\bibnamefont {Metlushko}}, \ and\ \bibinfo
  {author} {\bibfnamefont {L.}~\bibnamefont {Leonyuk}},\ }\href@noop {}
  {\bibfield  {journal} {\bibinfo  {journal} {Physica B: Condensed Matter}\
  }\textbf {\bibinfo {volume} {169}},\ \bibinfo {pages} {609} (\bibinfo {year}
  {1991})}\BibitemShut {NoStop}%
\bibitem [{\citenamefont {Pardo}\ \emph {et~al.}(1998)\citenamefont {Pardo},
  \citenamefont {de~la Cruz}, \citenamefont {Gammel}, \citenamefont {Bucher},\
  and\ \citenamefont {Bishop}}]{pardo_observation_1998}%
  \BibitemOpen
  \bibfield  {author} {\bibinfo {author} {\bibfnamefont {F.}~\bibnamefont
  {Pardo}}, \bibinfo {author} {\bibfnamefont {F.}~\bibnamefont {de~la Cruz}},
  \bibinfo {author} {\bibfnamefont {P.~L.}\ \bibnamefont {Gammel}}, \bibinfo
  {author} {\bibfnamefont {E.}~\bibnamefont {Bucher}}, \ and\ \bibinfo {author}
  {\bibfnamefont {D.~J.}\ \bibnamefont {Bishop}},\ }\href@noop {} {\bibfield
  {journal} {\bibinfo  {journal} {Nature}\ }\textbf {\bibinfo {volume} {396}},\
  \bibinfo {pages} {348} (\bibinfo {year} {1998})}\BibitemShut {NoStop}%
\bibitem [{\citenamefont {Troyanovski}\ \emph {et~al.}(1999)\citenamefont
  {Troyanovski}, \citenamefont {Aarts},\ and\ \citenamefont
  {Kes}}]{troyanovski_collective_1999}%
  \BibitemOpen
  \bibfield  {author} {\bibinfo {author} {\bibfnamefont {A.~M.}\ \bibnamefont
  {Troyanovski}}, \bibinfo {author} {\bibfnamefont {J.}~\bibnamefont {Aarts}},
  \ and\ \bibinfo {author} {\bibfnamefont {P.~H.}\ \bibnamefont {Kes}},\
  }\href@noop {} {\bibfield  {journal} {\bibinfo  {journal} {Nature}\ }\textbf
  {\bibinfo {volume} {399}},\ \bibinfo {pages} {665} (\bibinfo {year}
  {1999})}\BibitemShut {NoStop}%
\bibitem [{\citenamefont {Deng}\ \emph {et~al.}(2012)\citenamefont {Deng},
  \citenamefont {Tsuzuki}, \citenamefont {Miki}, \citenamefont {Felder},
  \citenamefont {Hara},\ and\ \citenamefont {Izumi}}]{deng_relaxation_2012}%
  \BibitemOpen
  \bibfield  {author} {\bibinfo {author} {\bibfnamefont {Z.}~\bibnamefont
  {Deng}}, \bibinfo {author} {\bibfnamefont {K.}~\bibnamefont {Tsuzuki}},
  \bibinfo {author} {\bibfnamefont {M.}~\bibnamefont {Miki}}, \bibinfo {author}
  {\bibfnamefont {B.}~\bibnamefont {Felder}}, \bibinfo {author} {\bibfnamefont
  {S.}~\bibnamefont {Hara}}, \ and\ \bibinfo {author} {\bibfnamefont
  {M.}~\bibnamefont {Izumi}},\ }\href@noop {} {\bibfield  {journal} {\bibinfo
  {journal} {J. Supercond. Novel Magn.}\ }\textbf {\bibinfo {volume} {25}},\
  \bibinfo {pages} {331} (\bibinfo {year} {2012})}\BibitemShut {NoStop}%
\bibitem [{\citenamefont {Monarkha}\ \emph {et~al.}(2012)\citenamefont
  {Monarkha}, \citenamefont {Timofeev},\ and\ \citenamefont
  {Shablo}}]{monarkha_violation_2012}%
  \BibitemOpen
  \bibfield  {author} {\bibinfo {author} {\bibfnamefont {V.~Y.}\ \bibnamefont
  {Monarkha}}, \bibinfo {author} {\bibfnamefont {V.~P.}\ \bibnamefont
  {Timofeev}}, \ and\ \bibinfo {author} {\bibfnamefont {A.~A.}\ \bibnamefont
  {Shablo}},\ }\href@noop {} {\bibfield  {journal} {\bibinfo  {journal} {Low
  Temp. Phys.}\ }\textbf {\bibinfo {volume} {38}},\ \bibinfo {pages} {31}
  (\bibinfo {year} {2012})}\BibitemShut {NoStop}%
\bibitem [{\citenamefont {Feigel知an}\ \emph {et~al.}(1989)\citenamefont
  {Feigel知an}, \citenamefont {Geshkenbein}, \citenamefont {Larkin},\ and\
  \citenamefont {Vinokur}}]{feigelman_theory_1989}%
  \BibitemOpen
  \bibfield  {author} {\bibinfo {author} {\bibfnamefont {M.~V.}\ \bibnamefont
  {Feigel知an}}, \bibinfo {author} {\bibfnamefont {V.~B.}\ \bibnamefont
  {Geshkenbein}}, \bibinfo {author} {\bibfnamefont {A.~I.}\ \bibnamefont
  {Larkin}}, \ and\ \bibinfo {author} {\bibfnamefont {V.~M.}\ \bibnamefont
  {Vinokur}},\ }\href@noop {} {\bibfield  {journal} {\bibinfo  {journal} {Phys.
  Rev. Lett.}\ }\textbf {\bibinfo {volume} {63}},\ \bibinfo {pages} {2303}
  (\bibinfo {year} {1989})}\BibitemShut {NoStop}%
\bibitem [{\citenamefont {Vinokur}\ \emph {et~al.}(1991)\citenamefont
  {Vinokur}, \citenamefont {Feigel知an},\ and\ \citenamefont
  {Geshkenbein}}]{vinokur_exact_1991}%
  \BibitemOpen
  \bibfield  {author} {\bibinfo {author} {\bibfnamefont {V.~M.}\ \bibnamefont
  {Vinokur}}, \bibinfo {author} {\bibfnamefont {M.~V.}\ \bibnamefont
  {Feigel知an}}, \ and\ \bibinfo {author} {\bibfnamefont {V.~B.}\ \bibnamefont
  {Geshkenbein}},\ }\href@noop {} {\bibfield  {journal} {\bibinfo  {journal}
  {Phys. Rev. Lett.}\ }\textbf {\bibinfo {volume} {67}},\ \bibinfo {pages}
  {915} (\bibinfo {year} {1991})}\BibitemShut {NoStop}%
\bibitem [{\citenamefont {Abulafia}\ \emph {et~al.}(1995)\citenamefont
  {Abulafia}, \citenamefont {Shaulov}, \citenamefont {Wolfus}, \citenamefont
  {Prozorov}, \citenamefont {Burlachkov}, \citenamefont {Yeshurun},
  \citenamefont {Majer}, \citenamefont {Zeldov},\ and\ \citenamefont
  {Vinokur}}]{abulafia_local_1995}%
  \BibitemOpen
  \bibfield  {author} {\bibinfo {author} {\bibfnamefont {Y.}~\bibnamefont
  {Abulafia}}, \bibinfo {author} {\bibfnamefont {A.}~\bibnamefont {Shaulov}},
  \bibinfo {author} {\bibfnamefont {Y.}~\bibnamefont {Wolfus}}, \bibinfo
  {author} {\bibfnamefont {R.}~\bibnamefont {Prozorov}}, \bibinfo {author}
  {\bibfnamefont {L.}~\bibnamefont {Burlachkov}}, \bibinfo {author}
  {\bibfnamefont {Y.}~\bibnamefont {Yeshurun}}, \bibinfo {author}
  {\bibfnamefont {D.}~\bibnamefont {Majer}}, \bibinfo {author} {\bibfnamefont
  {E.}~\bibnamefont {Zeldov}}, \ and\ \bibinfo {author} {\bibfnamefont {V.~M.}\
  \bibnamefont {Vinokur}},\ }\href@noop {} {\bibfield  {journal} {\bibinfo
  {journal} {Phys. Rev. Lett.}\ }\textbf {\bibinfo {volume} {75}},\ \bibinfo
  {pages} {2404} (\bibinfo {year} {1995})}\BibitemShut {NoStop}%
\bibitem [{\citenamefont {Burlachkov}\ \emph {et~al.}(1998)\citenamefont
  {Burlachkov}, \citenamefont {Giller},\ and\ \citenamefont
  {Prozorov}}]{ProzorovPRB58-98}%
  \BibitemOpen
  \bibfield  {author} {\bibinfo {author} {\bibfnamefont {L.}~\bibnamefont
  {Burlachkov}}, \bibinfo {author} {\bibfnamefont {D.}~\bibnamefont {Giller}},
  \ and\ \bibinfo {author} {\bibfnamefont {R.}~\bibnamefont {Prozorov}},\
  }\href@noop {} {\bibfield  {journal} {\bibinfo  {journal} {Phys. Rev. B}\
  }\textbf {\bibinfo {volume} {58}},\ \bibinfo {pages} {15067} (\bibinfo {year}
  {1998})}\BibitemShut {NoStop}%
\bibitem [{\citenamefont {Gurevich}\ and\ \citenamefont
  {K{\"u}pfer}(1993)}]{gurevich_time_1993}%
  \BibitemOpen
  \bibfield  {author} {\bibinfo {author} {\bibfnamefont {A.}~\bibnamefont
  {Gurevich}}\ and\ \bibinfo {author} {\bibfnamefont {H.}~\bibnamefont
  {K{\"u}pfer}},\ }\href@noop {} {\bibfield  {journal} {\bibinfo  {journal}
  {Phys. Rev. B}\ }\textbf {\bibinfo {volume} {48}},\ \bibinfo {pages} {6477}
  (\bibinfo {year} {1993})}\BibitemShut {NoStop}%
\bibitem [{\citenamefont {Hor}\ \emph {et~al.}(2005)\citenamefont {Hor},
  \citenamefont {Welp}, \citenamefont {Ito}, \citenamefont {Xiao},
  \citenamefont {Patel}, \citenamefont {Mitchell}, \citenamefont {Kwok},\ and\
  \citenamefont {Crabtree}}]{HorCrabtreeAPL87-05}%
  \BibitemOpen
  \bibfield  {author} {\bibinfo {author} {\bibfnamefont {Y.~S.}\ \bibnamefont
  {Hor}}, \bibinfo {author} {\bibfnamefont {U.}~\bibnamefont {Welp}}, \bibinfo
  {author} {\bibfnamefont {Y.}~\bibnamefont {Ito}}, \bibinfo {author}
  {\bibfnamefont {Z.~L.}\ \bibnamefont {Xiao}}, \bibinfo {author}
  {\bibfnamefont {U.}~\bibnamefont {Patel}}, \bibinfo {author} {\bibfnamefont
  {J.~F.}\ \bibnamefont {Mitchell}}, \bibinfo {author} {\bibfnamefont {W.~K.}\
  \bibnamefont {Kwok}}, \ and\ \bibinfo {author} {\bibfnamefont {G.~W.}\
  \bibnamefont {Crabtree}},\ }\href@noop {} {\bibfield  {journal} {\bibinfo
  {journal} {Appl. Phys. Lett.}\ }\textbf {\bibinfo {volume} {87}},\ \bibinfo
  {pages} {142506} (\bibinfo {year} {2005})}\BibitemShut {NoStop}%
\bibitem [{\citenamefont {Pompeo}\ and\ \citenamefont
  {Silva}(2008)}]{pompeo_reliable_2008}%
  \BibitemOpen
  \bibfield  {author} {\bibinfo {author} {\bibfnamefont {N.}~\bibnamefont
  {Pompeo}}\ and\ \bibinfo {author} {\bibfnamefont {E.}~\bibnamefont {Silva}},\
  }\href@noop {} {\bibfield  {journal} {\bibinfo  {journal} {Phys. Rev. B}\
  }\textbf {\bibinfo {volume} {78}},\ \bibinfo {pages} {094503} (\bibinfo
  {year} {2008})}\BibitemShut {NoStop}%
\bibitem [{\citenamefont {Golosovsky}\ \emph {et~al.}(1996)\citenamefont
  {Golosovsky}, \citenamefont {Tsindlekht},\ and\ \citenamefont
  {Davidov}}]{GolosovskyDavidovSUPERCONDSCITECHNOL9-96}%
  \BibitemOpen
  \bibfield  {author} {\bibinfo {author} {\bibfnamefont {M.}~\bibnamefont
  {Golosovsky}}, \bibinfo {author} {\bibfnamefont {M.}~\bibnamefont
  {Tsindlekht}}, \ and\ \bibinfo {author} {\bibfnamefont {D.}~\bibnamefont
  {Davidov}},\ }\href@noop {} {\bibfield  {journal} {\bibinfo  {journal}
  {Supercond. Sci. Technology}\ }\textbf {\bibinfo {volume} {9}},\ \bibinfo
  {pages} {1} (\bibinfo {year} {1996})}\BibitemShut {NoStop}%
\bibitem [{\citenamefont {Landau}\ and\ \citenamefont
  {Lifshitz}(1987)}]{LandauVol6FluidMechanics}%
  \BibitemOpen
  \bibfield  {author} {\bibinfo {author} {\bibfnamefont {L.~D.}\ \bibnamefont
  {Landau}}\ and\ \bibinfo {author} {\bibfnamefont {E.~M.}\ \bibnamefont
  {Lifshitz}},\ }\href@noop {} {\emph {\bibinfo {title} {Fluid Mechanics}}},\
  \bibinfo {series} {Course of Theoretical Physics}, Vol.~\bibinfo {volume}
  {6}\ (\bibinfo  {publisher} {Butterworth-Heinemann},\ \bibinfo {year}
  {1987})\BibitemShut {NoStop}%
\bibitem [{\citenamefont {Bass}\ \emph {et~al.}(1998)\citenamefont {Bass},
  \citenamefont {Shapiro}, \citenamefont {Shapiro},\ and\ \citenamefont
  {Shvartser}}]{bass_nonlinear_1998}%
  \BibitemOpen
  \bibfield  {author} {\bibinfo {author} {\bibfnamefont {F.}~\bibnamefont
  {Bass}}, \bibinfo {author} {\bibfnamefont {B.}~\bibnamefont {Shapiro}},
  \bibinfo {author} {\bibfnamefont {I.}~\bibnamefont {Shapiro}}, \ and\
  \bibinfo {author} {\bibfnamefont {M.}~\bibnamefont {Shvartser}},\ }\href@noop
  {} {\bibfield  {journal} {\bibinfo  {journal} {Physica C}\ }\textbf {\bibinfo
  {volume} {297}},\ \bibinfo {pages} {269} (\bibinfo {year}
  {1998})}\BibitemShut {NoStop}%
\bibitem [{\citenamefont {{Bryksin}}\ and\ \citenamefont
  {{Dorogovtsev}}(1993)}]{bryksin_nonlinear_1993}%
  \BibitemOpen
  \bibfield  {author} {\bibinfo {author} {\bibfnamefont {V.~V.}\ \bibnamefont
  {{Bryksin}}}\ and\ \bibinfo {author} {\bibfnamefont {S.~N.}\ \bibnamefont
  {{Dorogovtsev}}},\ }\href@noop {} {\bibfield  {journal} {\bibinfo  {journal}
  {Sov. J. Exp. Theor. Phys.}\ }\textbf {\bibinfo {volume} {77}},\ \bibinfo
  {pages} {791} (\bibinfo {year} {1993})}\BibitemShut {NoStop}%
\bibitem [{\citenamefont {Bondarenko}\ \emph {et~al.}(2006)\citenamefont
  {Bondarenko}, \citenamefont {Shablo},\ and\ \citenamefont
  {Koverya}}]{bondarenko_study_2006}%
  \BibitemOpen
  \bibfield  {author} {\bibinfo {author} {\bibfnamefont {S.~I.}\ \bibnamefont
  {Bondarenko}}, \bibinfo {author} {\bibfnamefont {A.~A.}\ \bibnamefont
  {Shablo}}, \ and\ \bibinfo {author} {\bibfnamefont {V.~P.}\ \bibnamefont
  {Koverya}},\ }\href@noop {} {\bibfield  {journal} {\bibinfo  {journal} {Low
  Temp. Phys.}\ }\textbf {\bibinfo {volume} {32}},\ \bibinfo {pages} {628}
  (\bibinfo {year} {2006})}\BibitemShut {NoStop}%
\end{thebibliography}%

\end{document}